\documentclass[%
 reprint,
 amsmath,amssymb,
 aps,
]{revtex4-1}

\usepackage{graphicx}% Include figure files
\usepackage{dcolumn}% Align table columns on decimal point
\usepackage{bm}% bold math
\usepackage{enumitem}
\usepackage{lipsum}
\usepackage{graphicx,epsfig,latexsym,amssymb}
\usepackage{multirow,amsmath,array,booktabs}
\usepackage{dcolumn}
\usepackage[section]{placeins}
\usepackage{epstopdf}
\usepackage{graphicx}
\usepackage{float}
\usepackage{slashed}
\usepackage{mathtools}

\DeclarePairedDelimiterX\braket[2]{\langle}{\rangle}{#1 \delimsize\vert #2}
\DeclarePairedDelimiterX\inner[2]{\langle}{\rangle}{#1,#2}

\begin{document}

\preprint{APS/123-QED}

\title{Properties of Low-Lying Charmonium States in a  Phenomenological Approach}
\preprint{APS/123-QED}

\author{Praveen P D'Souza $^{1,2}$}

\author{Antony Prakash Monteiro $^{1}$}%
 \email{aprakashmonteiro@gmail.com}
\affiliation{%
 $^1$P. G. Department of Physics, St Philomena college
              Darbe, Puttur  574 202, India
}%

%\author{Manjunath Bhat $^{2}$}%

\author{K. B. Vijaya Kumar $^{2}$}
\affiliation{
 $^2$Department of Physics, Mangalore University,
Mangalagangothri P.O., Mangalore - 574199, INDIA\\
%$^3$Department of Physics, Mangalore University,
%Mangalagangothri P.O., Mangalore - 574199, INDIA
}%}%

\date{\today}% It is always \today, today,

\begin{abstract}
We investigate the spectrum and decay rates of low lying charmonium states within the framework of the non relativistic quark model by employing a Coulomb like potential  from the  perturbative one gluon exchange and the linear confining potential along with the potential derived from instanton vacuum to account for the hyperfine mass splitting of charmonium states in variational approach. We predict radiative E1, M1, two-photon, leptonic and two-gluon decay rates of low lying charmonium states. An overall agreement is obtained with the experimental masses and decay widths.

\end{abstract}

\pacs{Valid PACS appear here}% PACS, the Physics and Astronomy
\maketitle

%\tableofcontents

\section{INTRODUCTION}

Charmonia are bound states of a charm and an anticharm quark $(c\bar{c})$, and represent an important testing ground for the properties of the strong interaction. There has been a great progress in the observation of the charmonium states from  the past few years.  The discovery of the first charmonium state $J/\psi$\cite{AJ74,AJ74b} has revolutionized the field of hadron spectroscopy. This led to a clear understanding of the prevailing theory of particle physics. Several quarkonium states have been observed after the discovery of the charmonium state $J/\psi$ at BNL and SLAC\citep{PDG}. The first observation of a singlet ground state of charmonium $\eta_c$ was done by Mark II and Crystal Ball experiments in the radiative decays of $J/\psi$ and $\psi'$\citep{PDG}. The discoveries of conventional states $h_c(1P)$, $h_c(2P)$, $\chi_c(1P)$, $\chi_c(2P)$, $\eta_c(1S)$ and the observation of the exotic states like X(3872), X(3915), Y(4260), Z(3930) at Belle, BaBar, LHC, BESIII,CLEO, etc  have created a renewed interest in quarkonium physics\cite{PDG}.  These new observations have given a deeper understanding of the  charmonium physics and have unraveled many mysteries\cite{WD16}.  Charmonium system is a powerful tool for the study of forces between quarks in QCD  in non-perturbative regime. Studies of charmonia production can improve our understanding of heavy quark production and the formation of bound states.

 The exploration and understanding of the substructure of hadrons, presented in terms of quarks and gluons by quantum chromodynamics(QCD),  has led to a  considerable progress in the study of charmonium states. Though there have been numerous calculations of charmonium spectra from  first principles  such as Lattice QCD\cite{CT95,JD08} and NRQCD which provide  rigorous theoretical implications for the experimental observations,quark model calculations  provide more intuitive insights  and supply important phenomenological guidance towards their study. The quark antiquark potential cannot be obtained from the first principles of QCD. Therefore, one has to use potential models to explain the observed hadronic properties. The QCD inspired potential models have been playing an important role in investigating heavy quarkonium. Most of the quark potential models \cite{EE76,EE78,EE80,SN85,SD80,ED03,GS95,FL91,FL99,FL94,GS94,GS96} have common ingredients in the non relativistic quark models(NRQM). Recently it is shown that the NRQM with instanton induced interaction(III)  explains $^1S_0$  and $^3S_1$ the nucleon- nucleon potential\cite{CS2017}.
 
  The NRQM formalism  provides a  systematic treatment of the perturbative and non perturbative components of QCD at hadronic scale \cite{BC81,BG97}. These QCD inspired potential models use a short range part motivated by perturbative QCD (Coulomb like or one gluon exchange potential) \cite{BC81,SG88,YY92,VA95,MG00,WJ91,KB13,FP1,FP2,FP3} and a phenomenological long range part accounting for confinement (i.e. linear, logarithmic or quadratic potential)\cite{MB17,MB177,KB97,APM11,BG12,BKB11,BKB13,BV11}. Our Model uses the NRQM formalism for the study of low lying charmonium states using a Hamiltonian which has the heavy quark potential from the instanton vacuum depending on r, the inter quark distance. The heavy quark potential derived from the instanton ensemble rises linearly as the relative distance between the quark and antiquark increases, then it  gets saturated. As the quark and the antiquark distance  increases the central  potential turns out to be  Coulomb like potential. Therefore to study the mass spectra of the quarkonia we have added Coulombic type  potential to central instanton potential.  This can be understood as a non perturbative contribution to the perturbative  potential from  instanton vacuum at large inter quark distances greater than the instanton size. At large distance  the instanton vacuum cannot explain quark confinement, hence  in our phenomenological potential model we have included  a linear confinement potential. Taking into  consideration the above factors, we have developed a nonrelativistic potential model  to obtain a reliable estimate of the masses of the low lying charmonium states and their decay widths.

The paper is organized in 4 sections. In sec.~\ref{intro} we briefly review the theoretical background for the non relativistic model.  In sec.~\ref{sec:RD} we present the results and discussions. In sec.~\ref{sec:C} we draw up conclusions. 
\section{Theoretical Background }
\label{intro}
\subsection{The Model}
In a potential model approach the entire dynamics of quarks in a meson is governed by a Hamiltonian which is composed of a kinetic energy term $K$ and a potential energy term V, that is, 
\begin{equation}
H=K+V.
\end{equation}
The kinetic energy  K is given by,
\begin{equation}
K= M+\frac{p^2}{2\mu}
\end{equation}  
Here p is the relative momentum, $\mu = \frac{m_Q m_{\bar{Q}}}{m_Q +m_{\bar{Q}}}$ is the reduced mass of the  $Q\bar{Q}$  system,  where $m_Q$ and $m_{\bar{Q}}$  are the masses of the individual quark and 
anti quark respectively and M is the total mass of quark and antiquark\cite{BGH}.\\

The potential energy  V is the sum of the heavy-quark potential $V_{Q\bar{Q}}(\vec{r})$,confining potential $V_{conf}(\vec{r})$ and Coulomb potential $V_{coul}(\vec{r})$,that is
\begin{equation}
V(\vec{r})=V_{Q\bar{Q}}(\vec{r})+V_{coul}(\vec{r})+V_{conf}(\vec{r})
\end{equation}
The heavy-quark potential $V_{Q\bar{Q}}(\vec{r})$ is, 
\begin{equation}
V_{Q\bar{Q}}(\vec{r}) =V_C(\vec{r})+V_{SD}(\vec{r}).
\end{equation}
Here $V_C(\vec{r})$ and $V_{SD}(\vec{r})$ are central  and spin dependent potentials due to instanton vacuum  respectively\cite{YU}.\\
$V_{C}(\vec{r})$ is given by the following expression
\begin{equation}
\label{eq4}
V_{C}(\vec{r}){\simeq}\frac{4\pi\bar{\rho}^3}{\bar{R^4}N_c}\Bigg(1.345\frac{r^2}{\bar{\rho}^2}-0.501\frac{r^4}{\bar{\rho}^4}\Bigg)
\end{equation}
 Here, $\rho=\frac{1}{3}$ fm the average size of the instanton, $ \bar{R}= 1 $ fm the  average separation between instantons and number of colors $ N_C$ is 3.

The spin- spin interaction $V_{SS}(\vec{r})$, the spin-orbit coupling term $V_{LS}(\vec{r})$ and the tensor part $V_T(\vec{r})$ contribute to the spin dependent potential;

\begin{equation}
\label{spin}
\begin{split}
V_{SS}(\vec{r})=\frac{1}{{3m_Q}^2}\nabla^2V_{C}(\vec{r});~
V_{LS}(\vec{r})=\frac{1}{{2m_Q}^2}\frac{1}{r}\frac{dV_{C}(\vec{r})}{dr};~\\
V_{T}(\vec{r})=\frac{1}{{3m_Q}^2}\Bigg(\frac{1}{r}\frac{dV_{C}(\vec{r})}{dr}-\frac{d^2V_{C}(\vec{r})}{dr^2}\Bigg).
\end{split}
\end{equation}

The coulomb-like (perturbative) one gluon exchange part of the potential is given by 
\begin{equation}
V_{coul}(\vec{r})=\frac{-4\alpha_s}{3r}
\end{equation}
with the strong coupling constant $\alpha_s$ and inter quark distance r.

The confinement term represents the non perturbative effect of QCD which includes the spin-independent linear confinement term\cite{BGH}
\begin{equation}
 V_{conf}(\vec{r})=-\Bigg[\frac{3}{4}V_0+\frac{3}{4}cr\Bigg]F_1{\cdot}F_2
\end{equation}
where c and $V_0$ are constants. F is related to the Gell-Mann matrix, $F_1 = \frac{ \lambda_1}{2} $ and $F_2 = \frac{ \lambda_2^*}{2} $ and
$F_1{\cdot}F_2 = \frac{-4}{3}$ for the mesons.\\

\subsection{Radiative Transitions}
The study of radiative transitions can help in understanding the  theory of strong interaction in the nonperturbative regime of QCD. The non-relativistic treatment adopted for the study of charmonium systems allows us to apply the usual multi-pole expansion in electrodynamics to compute the transition between the quarkonia states with the emission of a photon. The lowest order of multi-pole expansion tends to dominate the transition. The resulting transitions are the magnetic dipole $M_1$ and electric dipole $E_1$ transitions. The electric and magnetic dipole operators are decomposed by a multi-pole expansion into components with definite spherical tensor ranks.

In the phenomenological potential  model  the  expression used for the radiative decay width  does not directly influence the  structure of the potential model Hamiltonian. Thus the transition probabilities  are influenced by the chosen Hamiltonian only through the chosen wave function.    In a $M_1$ transition only the spin of the quarkonium state is changed, while the parity and the orbital angular momentum remain unchanged. The expression for the decay width of a spin flip $M_1$ transition between heavy quarkonium states depend on the radial matrix. The different radial matrix  elements can be obtained from the corresponding spin-flip magnetic moment operator.  The $M_1$ decay widths of charmonium are calculated using an assumption that the effective confining interaction is purely scalar.   In electric dipole $E_1$ transition, the parity of the states changes while spin remains unchanged. The one gluon exchange contribution survives in the $E_1$ transition, whereas the contribution from the scalar confining interaction term vanishes.

In calculating the radiative decay widths, we have assumed that in the non relativistic limit, the dipole radial matrix elements are independent of J, i.e all states within the same angular momentum multiplet have the same wave function \cite{EMD}. Radiative transitions could play an important role in the discovery and identification of charmonium states. They are sensitive to the internal structure of states, in particular to $^3L_L-^1L_L$  mixing for states with J = L. For our study of low lying charmonium states we have taken $E_1$ and $M_1$ radiative transitions since the other order transitions contribute a little to the radiative decays. The partial width for an $E_1$ radiative transition is given by, 
\begin{equation}
\Gamma(i\rightarrow f + \gamma)=\frac{4\alpha e_c^2}{3} (2J_f+1) S_{if}^E k^3_0\left|{\cal E}_{if}\right|^2 
\end{equation}
where $k_0=m_i-m_f$ is the energy of the emitted photon, $\alpha$ is the fine structure constant. $e_c=2/3$ is the charge of the c quark in units of $|e|$,  $m_i$ and $m_f$ are the masses of initial and final mesons,  $S_{if}^E=\rm{max}(L_i,L_f)\left \{\begin{array}{ccc}
J_i & 1 & J_f \\
L_f & S  & L_i 
 \end{array} \right\}^2$ is the statistical factor, $J_i$ and $J_f$ are the total angular momenta of initial and final mesons, $L_i$ and $L_f$ are the orbital angular momenta of initial and final mesons and $S$ is the spin of the initial meson. The radial overlap integral which has the dimension of length is, 
\begin{equation}
 {\cal E}_{if}=\frac{3}{k_0}\int^\infty_0 r^3R_{nl}(r)R'_{nl}(r)dr\left[\frac{k_0r}{2}j_0\left(\frac{k_0r}{2}\right)-j_1\left(\frac{k_0r}{2}\right)\right]
\end{equation}
with  $R_{nl}(r)$ and $R'_{nl}(r)$ as the normalized radial wave functions for the corresponding states and  $j_0$ and $j_1$ are spherical Bessel functions.\\

The $M_1$ transitions between S-wave $c\bar{c}$ states are given in the non relativistic approximation by \cite{WJ88,NL78,CE05,WE86,BB95,NB99,BA96},
\begin{equation}
\label{eqm1}
\Gamma(i\rightarrow f + \gamma)=\frac{4\alpha e_c^2}{3m_c^2} \frac{2J_f+1}{2L_i+1}\delta_{L_iL_f}\delta_{S_iS_f} k^3_0|M_{if}(r)|^2
\end{equation}
Here $M_{if}$ is the radial overlap integral which has the dimension of length, 
\begin{equation}
M_{if}= \int^{\infty}_04 \pi r^3 R_{nl}(r)j_0(kr/2)R'_{nl}(r) dr.
\end{equation}
In the overlap integral for unit operator between the coordinate wave functions of the initial and the final meson states, $j_0(kr/2)$ is the spherical Bessel function, $m_c$ is the mass of charm quark, $J_f$ is the total angular momentum of final meson state, $L_i$ is the orbital angular momentum of the initial meson state. 
\subsection{Annihilation Decays}
The annihilation decays of  charmonium states into gluons and light quarks make significant contributions to the total decay width of the states. The annihilation decays allow us to determine wave function at the origin. The annihilation decays of some $c\bar{c}$ states into  photons can be used as a tool for the production and identification of the resonances.

\subsubsection{Two Photon Decays}
 Two-photon  branching fraction for the charmonium provides a probe for the strong coupling constant at the charmonium scale via the two-photon decay width. This can be utilized as a sensitive test for the corrections for the non-relativistic approximation in the potential models  or  in the  effective  field  theories  such  as  non relativistic QCD (NRQCD).  The two-photon decays of P-wave charmonia are helpful for better understanding the nature of inter-quark forces and decay mechanisms. 

The $q\bar{q}$  quark pair in charge conjugation even states with $J\neq 1$ can annihilate into two photons. The expressions for the decay  rates of $n~^1S_0$, $n~^3P_0$ and $n~^3P_2$ states into two photons with the first order QCD radiative corrections are given  by \cite{KW88}.
\begin{eqnarray}
\Gamma(n~^1S_0\to \gamma\gamma)=\frac{3e^4_c\alpha^2}{m^2_c}|R_{nS}(0)|^2\left(1-\frac{3.4\alpha_s}{\pi}\right)\\
\Gamma(n~^3P_0\to \gamma\gamma)=\frac{27e^4_c\alpha^2}{m^4_c}|R'_{nP}(0)|^2\left(1+\frac{0.2\alpha_s}{\pi}\right)\\
\Gamma(n~^3P_2\to \gamma\gamma)=\frac{36e^4_c\alpha^2}{5m^4_c}|R'_{nP}(0)|^2\left(1-\frac{16\alpha_s}{\pi}\right)
\end{eqnarray}
The two photon decay widths of P wave charmonium states depend on the derivative of the radial wave function at the origin.
\subsubsection{Two Gluon Decays}
The even states  in charge conjugation of quarkonium with $J\neq 1$ can annihilate into two gluons, much in the same way as they decay into two photons. The charmonium states $^1S_0$, $^3P_0$, $^3P_2$ and $^1D_2$ can decay into two gluons, which account for a substantial portion of the hadronic decays for states below $c\bar{c}$ threshold. The two gluon decay widths are given by \cite{RV67,KW88}.
\begin{equation}
\Gamma(n~^1S_0\to 2g)=\frac{2\alpha^2_s}{3 m^2_c}|R_{nS}(0)|^2\left(1+\frac{4.8\alpha_s}{\pi}\right)
\end{equation}
\begin{equation}
\Gamma(n~^3P_0\to 2g)=\frac{6\alpha^2_s}{m^4_c}|R'_{nP}(0)|^2\left(1+\frac{9.5\alpha_s}{\pi}\right)
\end{equation}
\begin{equation}
\Gamma(n~^3P_2\to 2g)=\frac{8\alpha^2_s}{5m^4_c}|R'_{nP}(0)|^2\left(1-\frac{2.2\alpha_s}{\pi}\right)
\end{equation}
\begin{equation}
\Gamma(n~^1D_2\to 2g)=\frac{2\alpha^2_s}{3\pi m^6_c}|R''_{nD}(0)|^2
\end{equation}

It is natural that in the non-relativistic potential model of charmonium, the ratio of the two-photon and two-gluon  widths of the charmonium decays does not depend on the wave function and slowly grows with increase of the charmonium mass because of the proportionality to $\frac{1}{\alpha_s^2}$. According to QCD, the decay of charmonium is due to the annihilation of $c\bar{c}$ pair. The mass of $c\bar{c}$ pair is large and the annihilations of  $c\bar{c}$ into gluons are perturbative, so the two-gluon decay mode is dominant in the charmonium. The two gluon decay widths are sensitive to the behavior of the $q\bar{q}$ wave function and its derivatives near the origin.

\subsubsection{Leptonic Decays}
The vector mesons decay leptonically through interaction with the electromagnetic current. The  leptonic decay width is proportional to the average  value  of  the  squared  charge and the  squared  wave  function  at  the  origin.  Which  gives  the  probability  that  the quark and antiquark will interact with the photon at the origin of their relative coordinates and the mass of the vector mesons.   The  quark-antiquark assignments  for  the  vector  mesons,  as  well  as  the  fractional  values  for  the  quark  charges,  may  be  experimented  from  the values of their leptonic decay widths. 

The decay of vector meson into charged leptons proceeds through the virtual photon ($q\bar{q}\to l^+l^-$ where $l=e^-, \mu^-, \tau^-$). The $^3S_1$ and $^3D_1$ states have quantum numbers of a virtual photon, $J^{PC}=1^{--}$ and can annihilate into lepton pairs through one photon.

The leptonic decay width of the vector meson ($^3S_1$ charmonium) including the first order radiative QCD correction is given by \cite{RV67,KW88}
\begin{equation}
\Gamma(n~^3S_1\to e^+e^-)=\frac{4\alpha^2e^2_c|R_{nS}(0)|^2}{M^2_{nS}}\left(1-\frac{16\alpha_s}{3\pi}\right)
\end{equation}
where $\alpha\approx\frac{1}{137}$ is the fine structure constant, $M_{nS}$ is the mass of the decaying charmonium state and $e_c=2/3$ is the charge of the charm quark in units of the electron charge. For D wave charmonium states the leptonic decay width with leading order QCD correction is given by
\begin{equation}
\Gamma(n~^3D_1\to e^+e^-)=\frac{25\alpha^2e^2_c|R''_{nD}(0)|^2}{2m^4_cM^2_{nD}}\left(1-\frac{16\alpha_s}{3\pi}\right)
\end{equation} 
where $M_{nD}$ is the mass of the decaying charmonium state.  The leptonic  partial  widths  are  an  exploration  of  the  compactness  of  the  quarkonium  system  and  provide  important  information supplementary to level spacings. The quark-antiquark assignments for the vector mesons, as well as the  fractional values  for  the  quark  charges,  may  be  experimented  from  the  values  of  their  leptonic  decay  widths.
\section{Results and Discussions}
\label{sec:RD}
In our work, we have used the three-dimensional harmonic oscillator wave function which has been extensively used in atomic and nuclear physics is used as the trial wave function for obtaining the $Q\bar{Q}$ mass spectrum.  
\begin{equation}
\psi_{nlm}(r,\theta,\phi)=N\left(\frac{r}{b}\right)^l L_n^{l+1/2}(\frac{r}{b})exp\left(-\frac{r^2}{2b^2}\right)Y_{lm}(\theta,\phi)
\end{equation}
where $|N|$  is the normalizing constant given by 
\begin{equation}
|N|^2=\frac{2\alpha^3n!}{\sqrt{\pi}} \frac{2^{(2(n+l)+1)}}{(2n+2l+1)!}(n+l)!
\end{equation}
and $L_n^{l+1/2}(x)$ are the associated Laguerre polynomials,  
The harmonic oscillator wave function allows the separation of  motion of the center of mass and has been  used to study the spectra of baryons and mesons \cite{FD69,RP71}. If the basic states are  harmonic oscillator wave functions it is rather easy to evaluate the matrix elements of a few body systems such as mesons or baryons. In the harmonic oscillator wave function b is treated as a variational parameter, which is determined for each state by minimizing the expectation value of the Hamiltonian. The obtained b value is used in the harmonic oscillator wave function to find the mass spectrum \cite{BGH11}. 
We have two important parameters characterizing the dilute instanton liquid; the average size of the instanton $\rho=\frac{1}{3}$ fm, the values  of $\rho $ is less effective in the spin-dependent parts of the potential. The  average separation between instantons is $ \bar{R}= 1$ fm \cite{INSTB, INSTB2}. The strength of each part of the potential becomes stronger when smaller value of $ \bar{R}$ is employed. The instanton density is given as $N/V\simeq (200 MeV)^4$ and number of colors $ N_C$ is 3. Other parameters in our potential model are, the coupling constant $\alpha_s$, the charm quark mass $m_c$, the confinement strength c and a constant $V_0$.  The confinement strength $c$ is fixed by the stability condition for variation of mass of the vector meson against the size parameter $b$.  The mass of the charm quark $m_c$ and constant  $V_0$ were fixed so as to reproduce the ground state masses.  We start with a set of reasonable values of $m_c$ and $V_0$.   We used the following set of parameters in our work.

\begin{equation*}
\begin{split}
m_c=1475~{\rm MeV};~~~~~ \alpha_s = 0.3\\
~~c=260{\rm MeV~fm^{-1}};~~V_0=-125~ \rm{MeV};
\end{split}
\end{equation*}

\begin{widetext}

It should be noted that, for harmonic oscillator wave function  $|\Psi(0)|^2 \propto \frac{1}{b^3}$, which is required to estimate the leptonic and two photon and two gluon decay widths. The $\alpha_s$ quoted by the latest PDG  is 0.1182(12)\citep{PDG16}. The value of $\alpha_s$ used in the present investigation is 0.3. 

\begin{table}
\caption{\label{tbm}  Mass spectrum (MeV).}
\begin{tabular*}{\textwidth}{@{\extracolsep{\fill}}lllllllllllllll@{}}
\hline
$n^{2S+1}L_J$ & Name & $J^{PC}$&Present Work&$M_{exp}$ &\cite{SN85}&\cite{EDD}&\cite{LEPT1}&\cite{CL12}&\cite{JS13}\\
&&&MeV&MeV&MeV&MeV&MeV&MeV&MeV\\
\hline
$1^1S_0$&${\eta}_c{(1S)}$&$0^{-+}$&2984&2983.6$\pm$0.7&2970&2981&2981.7&2990.4&2990\\
$2^1S_0$&${\eta}_c{(2S)}$&$0^{-+}$&3640&3639.2$\pm$0.11&3620&3635&3619.2&3646.5&3643\\
$3^1S_0$&${\eta}_c{(3S)}$&$0^{-+}$&4061&....&4060&3989&4052.5&4071.9&4054\\
$1^3S_1$&$J/\psi$&$1^{--}$&3097&3096.916$\pm$0.011&3100&3096&3096.92&3085.1&3096\\
$2^3S_1$&${\psi}{(2S)}$&$1^{--}$&3687&3686.108$\pm$0.018&3680&3685&3686.1&3682.1&3703\\
$3^3S_1$&${\psi}{(3S)}$&$1^{--}$&4039&4039$\pm$1&4100&4039&4102&4100.2&4097\\
\hline
$1^1P_1$&$h_c(1P)$&$1^{+-}$&3525&3525.38$\pm$0.11&3520&3525&3523.7&3514.6&3515\\
$2^1P_1$&$h_c(2P)$&$1^{+-}$&3927&....&3960&3926&3963.2&3944.6&3956\\
$3^1P_1$&$h_c(3P)$&$1^{+-}$&4337&....&....&4337&....&4333.9&4278\\
$1^3P_0$&$\chi_{c0}(1P)$&$0^{++}$&3414&3414.75$\pm$0.31&3440&3413&3415.2&3351.6&3452\\
$2^3P_0$&$\chi_{c0}(2P)$&$0^{++}$&3916&3915$\pm$3$\pm$2&3920&3870&3864.3&3835.7&3909\\
$3^3P_0$&$\chi_{c0}(3P)$&$0^{++}$&4303&....&....&4301&....&4216.7&4242\\
$1^3P_1$&$\chi_{c1}(1P)$&$1^{++}$&3510&3510.66$\pm$0.07&3510&3511&3510.6&3500.4&3452\\
$2^3P_1$&$\chi_{c1}(2P)$&$1^{++}$&3872&3872&3950&3906&3950.0&3933.5&3947\\
$3^3P_1$&${\chi}_{c1}{(3P)}$&$1^{++}$&4312&....&....&4319&....&4317.9&4272\\
$1^3P_2$&$\chi_{c2}(1P)$&$2^{++}$&3555&3556.20$\pm$0.09&3550&3555&3556.2&3551.4&3532\\
$2^3P_2$&$\chi_{c2}(2P)$&$2^{++}$&3929&3927.2$\pm$2.6&3980&3949&3992.3&3979.8&3969\\
$3^3P_2$&${\chi}_{c2}{(3P)}$&$2^{++}$&4042&....&4010&4041&....&4383.4&4043\\
\hline
$1^1D_2$&${\eta}_{c2}{(1D)}$&$2^{-+}$&3812&....&3840&3807&3822.3&3807.3&3812\\
$2^1D_2$&${\eta}_{c2}{(2D)}$&$2^{-+}$&4198&....&4210&4196&4196.9&4173.7&4166\\
$1^3D_1$&${\psi}_1{(1D)}$&$1^{--}$&3779&3778&3820&3783&3789.4&3785.3&3796\\
$2^3D_1$&${\psi}_1{(2D)}$&$1^{--}$&4192&4191$\pm$5&4190&4159&4159.2&4150.4&4153\\
$1^3D_2$&${\psi}_2{(1D)}$&$2^{--}$&3823&3823&3840&3795&3822.1&3807.7&3810\\
$2^3D_2$&${\psi}_2{(2D)}$&$2^{--}$&4195&....&4210&4190&4195.8&4173.7&4160\\
$1^3D_3$&${\psi}_2{(1D)}$&$3^{--}$&3845&....&....&....&3844.8&3814.6&....\\
$2^3D_3$&${\psi}_2{(2D)}$&$3^{--}$&4220&....&....&....&4218.9&4182.9&....\\
\hline
\end{tabular*}
\vspace{0mm}
\end{table}
\end{widetext}
Table \ref{tbm} lists the low lying charmonium states in comparison with experimental data and other theoretical models. Our predictions for the masses agree with PDG data within a few MeV. The model correctly reproduces the mass spectrum of charmonium states.The mass of singlet state $\eta_c(1S)$ is found to be $2984$ MeV which is in good agreement with the experimentally measured mass value $2983.6\pm 0.7$MeV \cite{PDG}. Lattice QCD calculations predict a mass of $2985(1)$ MeV\cite{KT15} for $\eta_c(1S)$ state. The mass of the spin triplet state $J/\psi(1S)$ calculated in our model is $3097 $ MeV which is in good agreement with the experimental value $3096.96\pm0.011$MeV\cite{PDG} and the values of other theoretical models \cite{SN85},\cite{EDD},\cite{LEPT1},\cite{CL12},\cite{JS13}. The lattice QCD calculations predict a mass of $3099\pm 1$ MeV for $J/\psi(1S)$ state. The masses of radially excited Charmonium state $\eta_c(2S)$ and its triplet partner $\psi(2S)$ calculated in our model are in good agreement with both experimental value and with other theoretical models. The lattice QCD calculations predict a slightly less mass value for $\eta_c(2S)$,  ($M(\eta_c(2S))=3612\pm 9 $ MeV) and for $\psi(2S)$($M_{\psi(2S)}=3653\pm 12 $ MeV)\cite{KT15}. It is clear from the table, that the mass of $\eta_c(3S)$ and its triplet partner calculated in our model is reasonably in good agreement with the experimental and with other model values. The lattice QCD calculations predict a mass of $4074\pm 20$ MeV for $\eta_c(3S)$ and $4099\pm 24 $ MeV for $\psi(3S)$\cite{KT15}.
 
The mass of spin-singlet P-wave charmonium $h_c(1P)$ is in good agreement with experiment and other theoretical models. However the lattice QCD predicts a mass of $3506\pm 6$ MeV for $h_c(1P)$ state which is slightly below  the experimental value\cite{KT15}. The spin triplet states $\chi_{cJ}$ are in good agreement with both experiment and other phenomenological models. The lattice QCD obtains the masses $20-25$ MeV smaller than the experimental values for these states\cite{KT15}. We have also predicted masses of low lying  $D$ wave states which is reasonably in good agreement with available experimental  data \cite{PDG16}and with other theoretical models \cite{LEPT1},\cite{CL12},\cite{JS13}

The spin-orbit and tensor potentials in eqn(\ref{spin}) are responsible for the splitting of the charmonium levels. The hyperfine splitting of 1S state obtained in our model $\Delta M(1^3S_1-1^1S_0)$ is 113 MeV. The hyperfine mass splitting calculated in our model agrees with the experimental value $\Delta M(1^3S_1-1^1S_0) =113.2 \pm 0.7$ MeV\cite{PDG} and lattice QCD results $\Delta M(1^3S_1-1^1S_0)$ = 114 $\pm$ 1MeV. The hyperfine mass splitting of 2S states $\Delta M(2^3S_1-2^1S_0)$= 47 MeV is in good agreement with the  $\Delta M(2^3S_1-2^1S_0) = 47 \pm 1 $ MeV\cite{PDG}. The lattice QCD predicts a 2S hyperfine  splittings of $57.9 \pm 2.0$ MeV \cite{MD12} which is slightly higher than the experimental value. For $1^3P_1$ - $1^3 P_0$ splitting, we obtain $\Delta M(1^3P_1-1^3P_0) =96 $ MeV which is good agreement with the experimental value $\Delta M(1^3P_1-1^1P_0) =95.5 \pm 0.8 $ MeV \cite{PDG}. The Lattice QCD calculations predict a 1P splitting of $ 68.4 \pm 5.0 + 11.8 - 3.0$ MeV\cite{OM02} which is reasonably in good agreement with our calculations. For $1^3P_2-1^3P_1$ splitting, we obtain $\Delta M (1^3P_2 -1^3P_1)$=45 MeV, which is in good agreement with the experimental data $\Delta M (1^3P_2 -1^3P_1)=45.7 \pm 0.2 $ MeV \cite{PDG}.  The lattice QCD calculations predict $\Delta M (1^3P_2 -1^3P_1)= 31.4 \pm 8.4 $ MeV \cite{OM02}, which is rather a low value compared to experimental value. 

The ratio between the two hyperfine structures, $\frac{\Delta M (1^3P_2 -1^3P_1)}{\Delta M (1^3P_1 -1^3P_0)} $  sheds light on the nature of the confinement. Our estimate of the ratio is $\frac{\Delta M (1^3P_2 -1^3P_1)}{\Delta M (1^3P_1 -1^3P_0)} $ =0.46, and that from the experiment is  0.48\cite{OM02,LW96}. Another interesting quantity is the P-state hyperfine splitting $\Delta M(1^1P_1-1^3P)$, where,
\begin{equation}  
M(1^3P)=\frac{5M(1^3P_2)+3M(1^3P_1)+M(1^3P_0)}{9}
\end{equation}
is the center of gravity of the P-wave system.      
The P-state hyperfine splitting should be  much  smaller than S-state hyperfine splitting, since the P state wave function is zero at the origin. Our estimate of the P-state hyperfine splitting is   $\Delta M(1^1P_1-1^3P)$ = 0.6 MeV and the experimental value is $\Delta M(1^1P_1-1^3P)$ = 0.9 MeV \cite{PDG}.  The lattice QCD calculations obtain  P-state hyperfine splitting of -1.4 MeV. 

The spin averaged masses  are defined by, 
\begin{equation}
M(n\bar{S})=\frac{3M(n^3S_1)+ M(n^1S_0)}{4}
\end{equation}
\begin{equation}
M(n\bar{P})=\frac{3M(n^1P_1)+5M(n^3P_2)+3M(n^3P_1)+M(n^3P_0)}{12}
\end{equation} 
with  n=1,2,3,... the radial quantum numbers.  The spin averaged masses calculated in our model are $M(1\bar{S})$=3068.75 MeV, $M(2\bar{S})$=3675.25 MeV, $M(1\bar{P})$=3524.5 MeV and $M(2\bar{P})$=3913.16 MeV and the experimental value of spin averaged masses are 2984.3, 3638.5.6 MeV, 3525.3 MeV and 3929 MeV  respectively \cite{OM02} \cite{JZ13}. The calculated spin averaged mass splittings  are listed in Table \ref{split}.
\begin{widetext}
\begin{table}
\caption{\label{split} Spin averaged mass splittings (MeV)}
\begin{tabular*}{\textwidth}
{@{\extracolsep{\fill}}llllllll@{}}
\hline
Mass Splittings & Present Work &Exp&Lattice QCD \cite{OM02}&GI\cite{TB05}&\cite{WJ16}&\cite{BQ09}\\
\hline
$M(1^3S_1-1^1S_0)$&113&$113.2\pm 0.7$&$114\pm 1$&113&114&118\\
$M(2^3S_1-2^1S_0)$&47&$47\pm 1$&$57.9\pm2$&53&44&50\\
$M(1^3P_2-1^3P_1)$&45&$45.7\pm 0.2$&$31.4\pm 8.4$&40&36&44\\
$M(1^3P_1-1^3P_0)$&96&$95.5\pm 0.8$&68.4&65&101&77\\
$M(1^1P_1-1^3P)$&0.6&0.9&-1.4&&&\\
$M(1^1P_1-1\bar{S})$&456.25&458.5&$448.8\pm29$&&&\\
$M(1^3P_0-1\bar{S})$&345.25&347.1&$419.8\pm47$&&&\\
$M(1^3P_1-1\bar{S})$&441.25&442.9&$448.8\pm34$&&&\\
$M(1^3P_2-1\bar{S})$&486.25&488.6&$448.8\pm29$&&&\\
$M(1\bar{P}-1\bar{S})$&455.75&457.9&$457.9$&&\\
$M(2\bar{P}-2\bar{S})$&237.91&....&$462\pm72$&&&\\
$M(2\bar{S}-1\bar{S})$&606.75&595&$671\pm21$&&&\\
$M(2\bar{P}-1\bar{P})$&388.66&....&$675\pm76$&&&\\
\hline
\end{tabular*}
\end{table}
\end{widetext}
Radiative decays of excited charmonium states are  powerful tool to study the internal structure of the mesons. The possible $E1$ decay modes have been listed in Table \ref{tabe1}. Most of the predictions for $E1$ transitions are in qualitative agreement with other theoretical models. However, there are some differences in the predictions  due to differences in phase space arising from different mass predictions and also from the wave function effects.  We find our results are compatible with other theoretical model values for most of the channels.

The M1 transition rates of charmonium states have been calculated using eqn(\ref{eqm1}). Allowed M1 transitions correspond to triplet-singlet transition between S-wave states and between P-state of the same n quantum number, while hindered M1 transitions are either triplet-singlet or singlet-triplet transitions between S-wave states of different quantum numbers. In order to calculate decay rates of hindered transitions, we need to include relativistic corrections, viz, modification of the nonrelativistic wave functions, relativistic modification of the electromagnetic transition operator, and finite-size corrections. In addition to these, there are additional corrections arising from the quark anomalous magnetic moment.

 Corrections to the wave function that give contribution to the transition amplitude are of two categories: (1) higher order potential corrections, which are distinguished as (a) the zero recoil effect and (b)recoil effects of the final state meson, and (2) color octet effects. The color octet  effects have not been  included in potential model formulation and are not considered so far in radiative transitions. The spherical Bessel function $j_0(k_0r/2)$, introduced in eqn(\ref{eqm1}), takes into account  the so called finite size effect(equivalently, re-summing the multipole-expanded  magnetic amplitude to all orders). For small value of $k_0$, $j_0(k_0r/2) \rightarrow 1$, the transitions with $n'=n$ have dominant contribution to the  matrix elements,though the corresponding partial decay widths are suppressed by smaller $k_0^3$ factors. For a large value of photon energy (k), transitions with $n \neq n'$ have dominant contribution to the matrix element, since $j_0(k_0r/2)$ becomes very small. M1 transition rates are very sensitive to hyperfine splittings of the levels due to the $k_0^3$ factor in  eqn(\ref{eqm1}).   The resulting M1 radiative transition rates of these states are presented in  Table \ref{tabm1}. The M1 transition rates calculated in our model agree well with the values predicted by other theoretical models.
 
Using the Van-Royen-Weisskopf  relation  we have calculated annihilation decay widths like leptonic decay widths, two -photon and two gluon decay widths with the inclusion of radiative corrections \cite{RV67}. The resulting leptonic decay widths are listed in Table V. Our predictions for leptonic decay widths have been compared with experiment and other theoretical models and are found to be in good agreement. The Tables VI and VII list the two photon and two gluon decay widths of charmonium states. The two photon  and  two-gluon decay widths for charmonium states reasonably agree with the available experimental data and with  other theoretical models.  
\begin{widetext}

\begin{table}
\label{tabe1}\caption{E1 Transition rates}
\begin{tabular*}{\textwidth}
{@{\extracolsep{\fill}}lllllllll@{}}
\hline
E1 Transition&k&Present Work $\Gamma$&$\Gamma_{Expt}$&\cite{TB05}$^a$&\cite{TB05}$^b$&\cite{CE05}&\cite{CL12}&\cite{LEPT1}\\
$i\rightarrow f$& (MeV)&($i\rightarrow f$)(keV)&(keV)&(keV)&(keV)&(keV)&(keV)&(keV)\\
\colrule
$1^3P_0\rightarrow 1^3S_1+\gamma$&317&114.74&119.5$\pm$8&152&114&120&97&139.3\\
$1^3P_1\rightarrow 1^3S_1+\gamma$&413&297.61&295.8$\pm$13&314&239&241&330&38.4\\
$1^3P_1\rightarrow 1^1S_0+\gamma$&526&486.26&&498&352&482&465&546.4\\
$1^3P_2\rightarrow 1^3S_1+\gamma$&458&350.58&384.2$\pm$16&424&313&315&421&319.4\\
$2^3S_1\rightarrow 1^3P_0+\gamma$&273&55.61&&63&26&47&34&25.2\\
$2^1S_0\rightarrow 1^1P_1+\gamma$&115&40.26&&49&36&35.1&72&17.4\\
$2^3S_1\rightarrow 1^3P_1+\gamma$&177&25.45&28.0$\pm$1.2&54&29&42.8&48&29.1\\
$2^3S_1\rightarrow 1^3P_2+\gamma$&132&23.9&26.6$\pm$1.1&38&24&30.1&43&26.5\\
$1^3D_1\rightarrow 1^3P_0+\gamma$&365&200.66&199.3$\pm$25&403&213&299&367&243.9\\
$1^3D_1\rightarrow 1^3P_1+\gamma$&269&75.34&79.2$\pm$16&125&77&99&146&104.9\\
$1^3D_1\rightarrow 1^3P_2+\gamma$&224&3.70&3.88&$<24.6$&4.9&3.3&6.8&1.9\\
$1^3D_2\rightarrow 1^3P_1+\gamma$&313&245.35&&307&268&313&321&256.7\\
$1^3D_2\rightarrow 1^3P_2+\gamma$&268&80.64&&64&66&69.5&79&61.8\\
$1^1D_2\rightarrow 1^1P_1+\gamma$&287&310.52&&339&344&389&398&\\
$1^3D_3\rightarrow 1^3P_2+\gamma$&290&246.79&&272&296&402&340\\
\botrule
\textsuperscript{a}\footnotesize{Non relativistic quark model}\\
\textsuperscript{a}\footnotesize{Relativistic quark model}
\end{tabular*}
\end{table}

\begin{table}
\label{tabm1}
\caption{M1 Transition rates}
\begin{tabular*}{\textwidth}
{@{\extracolsep{\fill}}llllllll@{}}
\hline
$(M_1)$Transition&k&Present Work $\Gamma$&$\Gamma_{Expt}$&\cite{TB05}$^c$ &\cite{TB05} $^d$&\cite{CL12}$^c$ &\cite{CL12}$^d$\\
$i\rightarrow f$& (MeV)&($i\rightarrow f$)(keV)&(keV)&(keV)&(keV)&(keV)&(keV)\\
\colrule   
$1^3S_1\rightarrow 1^1S_0+\gamma$&113&2.38&1.58$\pm$0.37&2.9&2.4&1.5&2.2\\
$2^3S_1\rightarrow 2^1S_0+\gamma$&47&0.17&0.143$\pm$0.027&0.21&0.17&3.1&3.8\\
$2^3S_1\rightarrow 1^1S_0+\gamma$&703&3.96&0.97$\pm$0.027&4.6&9.6&6.1&6.9\\
$2^3S_1\rightarrow 1^3S_1+\gamma$&590&5.30&&7.9&5.6&0.70&0.71\\
$3^3S_1\rightarrow 2^1S_0+\gamma$&399&1.12&&0.61&2.6&3.2&3.7\\
$3^1S_0\rightarrow 2^3S_1+\gamma$&374&1.42&&1.3&0.84&1.7&1.6\\
$3^1S_0\rightarrow 1^3S_1+\gamma$&964&6.51&&6.3&6.9&5.9&6.5\\
$2^1P_1\rightarrow 1^3P_2+\gamma$&372&0.23&&0.071&0.11&1.3&1.2\\
$2^1P_1\rightarrow 1^3P_1+\gamma$&417&0.18&&0.058&0.36&0.16&0.13\\
$2^1P_1\rightarrow 1^3P_0+\gamma$&513&1.31&&0.033&1.5&5.6&5.3\\
$2^3P_2\rightarrow 1^1P_1+\gamma$&404&1.12&&0.67&1.3&1.0&0.89\\
$2^3P_1\rightarrow 1^1P_1+\gamma$&347&0.15&&0.050&0.045&0.15&0.13\\
\botrule
\textsuperscript{c}\footnotesize{Non relativistic quark model}\\
\textsuperscript{d}\footnotesize{Relativistic quark model}
\end{tabular*}
\end{table}

\begin{table}
\label{tablll}\caption{Leptonic Decay widths (keV)}
\begin{tabular*}{\textwidth}
{@{\extracolsep{\fill}}lllllllll@{}}
\hline
State & Present Work $\Gamma_{l^+l^-}$ &Exp $\Gamma_{l^+l^-}$&\cite{BGH}&\cite{LEPT1}&\cite{LEPT2}\\
\colrule
$J/\psi$&4.65&$5.55\pm$0.14&3.589&4.28&12.13\\
$\psi(2S)$&2.25&$2.33\pm0.07$&1.440&2.25&5.03\\
$\psi(3S)$&0.98&$0.86\pm0.07$&0.975&1.66&3.48\\
$1^3D_1$&0.35&$0.242\pm0.030$&0.096&0.09&0.056\\
$2^3D_1$&0.68&$0.83\pm0.07$&0.112&0.16&0.096\\
\botrule
\end{tabular*}
\end{table}

\begin{table}
\label{tab2p}\caption{Two-Photon  Decay widths (keV)}
\begin{tabular*}{\textwidth}
{@{\extracolsep{\fill}}llllll@{}}
\hline
 State & Present Work  $\Gamma$ &Exp$\Gamma$&\cite{BGH}&\cite{TPHT1}&\cite{TPHT2}\\
\colrule
$\eta_c(1S)$&6.89&$7.2\pm0.7$&6.812&3.50&7.18\\
$\eta_c(2S)$&6.68&$7.0\pm3.5$&2.625&1.38&1.71\\
$\eta_c(3S)$&1.23&&1.760&0.94&1.21\\
$\chi_{c0}(1P)$&1.96&$2.36\pm0.35$&2.119&1.39&3.28\\
$\chi_{c0}(2P)$&1.02&&1.308&1.11&\\
$\chi_{c2}(1P)$&0.82&$0.66\pm0.07$&0.261&0.44&\\
$\chi_{c2}(2P)$&0.19&&0.168&0.48&\\
\botrule
\end{tabular*}
\end{table}

\begin{table}[H]
\label{tab2g}\caption{Two-Gluon  Decay widths (MeV)}
\begin{tabular*}{\textwidth}
{@{\extracolsep{\fill}}llllll@{}}
\hline
 State & Present Work $\Gamma$ &Exp&\cite{BGH}&\cite{TGG1}&\cite{TGG2}\\
\colrule
$\eta_c(1S)$&27.61&$28.6\pm2.2$&22.048&15.70&32.209\\
$\eta_c(2S)$&7.92&$14\pm7$&8.496&8.10&\\
$\eta_c(3S)$&5.67&&5.696&&\\
$\chi_{c0}(1P)$&9.67&$10.3\pm0.6$&6.114&4.68&10.467\\
$\chi_{c0}(2P)$&3.67&&3.775&&\\
$\chi_{c2}(1P)$&2.15&$1.97\pm0.11$&0.633&1.72&1.169\\
$\chi_{c2}(2P)$&0.59&&0.401&&\\
$1^1D_2^e$ &0.068&&0.014&&\\
$2^1D_2^e$&0.061&&0.012&&\\
\botrule
\textsuperscript{e}\footnotesize{Without QCD corrections}\\

\end{tabular*}
\end{table}

\end{widetext}
\section{Conclusions and Outlook}
\label{sec:C}
Basic aim of the present work is to develop a consistent model which could reproduce both the spectra and the decay widths with the same set of parameters and to investigate the effect of instanton potential on masses and the excited states of charmonium. From our analysis, we infer that the present model has the right prediction both for the mass spectrum and decay widths. Since different potentials can reproduce the same spectra the stringent test for any given model is the calculation of the  other observables like leptonic, the radiative, two-photon  and two-gluon decay widths  in addition to mass spectrum.  In our earlier work also, we had come to the similar conclusion while investigating light meson spectrum \cite{KBH} \cite{BGH05}. The differences in the prediction for the decay rates in various theoretical models can be attributed to the differences in mass predictions and wave function effects. From the study of mass spectra and decay properties of $c\bar{c}$ states  in a phenomenological approach we draw up the following conclusions.
\begin{enumerate}
\item Our calculations for the low lying charmonium states are in good agreement with experimental measurements
\item The hyperfine separations are directly related to the spin-spin interaction. The theoretical predictions of our model  are remarkably consistent with well established experimental data for the conventional charmonium states. Below the open charm threshold, our theoretical calculations  are in  agreement with lattice calculations and experimental results. 
\item The mass splittings between the radial excitations and the ground states also provide an important check on the validity of our potential model. 
\item The calculations demonstrate that the charmonium masses/mass splittings can be computed with a combination of coulombic, instanton and confinement potentials. Hence,in heavy quark sector,the instanton potential plays the role of OGEP. The instanton effects are quite significant in the central part of the potential.
 \item They turn out to be rather small in describing the hyperfine mass splittings of the charmonia. This might be due to the spin-dependent part of the potential from the instanton vacuum is  an order of magnitude smaller than the central part. The tensor interaction almost does not contribute to the masses.
 \item Charmonium decays provide a deeper insight on the exact nature of the inter quark forces and decay mechanisms. Using the predicted masses and the radial wave function at the origin, leptonic decays,two photon and two gluon decays are computed using the Van Royen-Weisskopf relation. The calculated values including the correction factor agree with the experimental values within a few MeV. From our calculations, we conclude that the inclusion of QCD correction factors are of importance for obtaining accurate results for  decay rates.
 
\end{enumerate}

%\begin{figure}
%\resizebox{0.75\textwidth}{!}{%
  %\includegraphics{leer.eps}
%\vspace{5cm}       % Give the correct figure height in cm
%\end{figure}
%\begin{figure*}
%\vspace*{5cm}       % Give the correct figure height in cm
%\end{figure*}
\bibliography{mybib}
\end{document}